\documentstyle[twocolumn,aps,prb,prbbib,psfig]{revtex} %% REV
\title{Growth Models And The Question Of Universality Classes } 
\author{W.~E.~Hagston, and H.~Ketterl $^{\ast }$}
\address{Department of Applied Physics, University of Hull,\\
HU6 7RX, U.K.}
\date{\today}

\input epsf
\begin{document}
\draft
\maketitle
\begin{abstract}
In the past many papers have appeared which simulated surface growth with different growth models.  The results showed that, if models differed 
only slightly in their `growth' rules, the resulting surfaces may belong to different universality classes, i.e. they are described by
different differential equations. In the present paper we describe a mapping of ``growth rules'' to differential operators and give
plausibility arguments for this mapping. We illustrate the validity of our theory by applying it to published results. 

\end{abstract}

%\newpage
\section{Introduction}
During the past few years the kinetic roughening of surfaces has become a field of increasing interest. In particular, many papers have appeared concerned with  
computer simulations of surface growth [see e.g. Ref. \onlinecite{barabasi}].

 In general, the surface is characterized by a height $h$ appropriate to a $d$
 dimensional substrate of size $L$. The width of the surface $w(t,L)$ at a time $t$ is
characterized by $w(t,L) = \sqrt{ \overline{ (h)^2} - \overline{(h)}^2  }$
where the bar denotes an average. If the resulting surface is self-affine it can
be represented by a dynamical scaling law

\begin{equation}
w(t,L) \sim L^{\alpha }f\left(\frac{t}{L^z}\right),
\label{e1}
\end{equation} 

where the function $f(x) \rightarrow $ constant for $x \rightarrow \infty $ and $ f(x) \sim x^{\beta },$ with $\beta = \alpha /z $ as
$x \rightarrow 0 $. The unit of time corresponds to depositing $L$
particles. For models which do not contain vacancies or overhangs, so-called solid on
solid models (SOS), this means that the average height $\overline{h} $ and the time $t$
are identical.  The exponents $\alpha ,~ \beta $ and $z$ determine which universality class the given model
belongs to. In the present paper we wish to examine the implications of some of the
standard assumptions in the theory, and in particular to provide a mapping of
prescribed rules of growth to the corresponding differential operators appearing in the
associated stochastic growth equations.

\section{Theory and Discussion}
\subsection{Simple growth systems}

The dynamical evolution of a surface prior to any movement of the deposited particles is presumed to be described by the equation 
\begin{equation}
\frac{\partial h(x,t)}{\partial t} = F + \eta (x,t)
\label{e2}
\end{equation}
\noindent where $\eta $ is a noise term with zero mean (i.e. $ \overline{\eta }=0$) and F is the flux rate of incident particles. For the particular case, usually considered in the literature, that
$F$ is a constant we obtain $\overline{h} = Ft$. 
The last result is only true if there are no vacancies in the system. 
 A similar result (but with $F$ replaced
by a larger entity, $F'$ say) would also hold, if the
vacancy concentration stays constant over time,  
i.e. in this case we would again have $\overline{h}=F't$. However, for such a situation the
starting equation would not be equal to Eq.(\ref{e2}), and $F' \neq F$, the incident
flux. In general, the form of the starting equation for situations where vacancies and
overhangs occur is far from obvious and is a problem we will return to later. Before
we can answer such questions it is pertinent to address a far simpler problem.  Namely,
what type of deposition process is Eq. (\ref{e2}) applicable to anyway? One class of such
processes is the so called SOS models in which particles are deposited randomly at
lattice sites,  i.e.\ a number is chosen at random  and the number of particles at the
site characterized by that number is increased by one. This is usually referred to as
(pure) random deposition (RD) and corresponds to a constant flux $F$ with a random noise
term $\eta $. (We refer to it as pure since the deposited atoms are not permitted to
move.) It is well known that the interface width for (pure)RD increases indefinitely with
time,  i.e.\ the associated surface is not self-affine with the consequence that the
width itself does not saturate. \cite{barabasi}

A variant of (pure) RD is to allow the particles to move after they have been deposited.
Two types of movement are possible. One is `horizontal' movement in which the height of
the moving particle does not change, and the other is vertical movement in which the
height of the particle does change. With regard to the latter there are two possible
types. One is upward vertical movement in which the height of the particle increases
and the other is downward vertical movement in which the height of the moving particle
decreases. In order to complete the rules of growth  with regard to the `allowed' moves we need to
specify the conditions under which horizontal or vertical movement is terminated. We
will refer to these collectively as `sticking rules'. 

In what follows we will show that what superficially appear to be `trivial' or 
`obvious' rules of movement (often designed for convenience of implementation on a
computer) can have extremely subtle implications for the resulting mathematics (and, in
particular, the associated differential equations purporting to model the resulting
surface structure). Furthermore, we will show that the implications for the mathematics
(even for the same set of rules) can be different dependent on the dimensionality
of the surface $d'$ under consideration.

To be definite, and also for reasons of simplicity, we will focus attention initially on
a $ d'=1$ surface i.e. a two dimensional structure $(d=2)$ having a $y$-axis (the height) and
an $x$-axis (the coordinate defining the position along the surface). Let us consider
first a variant on the pure RD model in which we allow both horizontal and vertical
motion (such motion being described under the general heading of surface relaxation
processes). Within the constraint of SOS, the allowed moves could be the following.
Permit the particle to be displaced horizontally up to a maximum of $l$ lattice
sites provided always that there is another particle immediately below it. If such a
situation always prevails, then at the end of the $l$ moves leave the particle where it
is. If, on the other hand, in the course of making these moves the particle encounters a
gap, then allow it to move vertically down until it reaches a position where there is
again a particle immediately below it  at which point the particle movement is
terminated. Similarly, if in the course of making up to $l$ moves, it meets another
particle on the same level all movement is terminated. 

It is clear that surface relaxation processes involving such SOS movements have the effect of changing the noise term $\eta $
in Eq (\ref{e2}) to a new value $\eta '$. Thus this situation could be described by the
equation 

\begin{equation}
\frac{\partial h}{\partial t} = F + \eta ' ~ .
\label{e3}
\end{equation}
Such an equation, although correct, is of little use from an analytical viewpoint
because we do not know the form of the noise term $\eta ' $. However the essential
characteristic of SOS movement that we can make use of is that it leaves the average
height of the surface unchanged. In effect one lattice site loses a particle and
another lattice site (which could be anywhere up to a distance $l$ away horizontally)
gains one. This suggests that we write the noise term $\eta ' $ in the form 
\begin{equation}
\eta ' = \eta + \frac{\partial G}{\partial x}
\label{e4}
\end {equation}
where $G$ is any (well behaved) function we care to choose. The reason being that the act
of averaging over a finite (but large) number of discrete particles is deemed as being
equivalent (in a continuum description) to the act of integrating and then dividing by the
`length' of the region of integration. Consequently, expressing the change in the noise
as a divergence means (at least in the limit that the length of the system becomes
infinite) that the average of $\eta '$ is zero (just like $\eta $). This then gives the
formal result 
\[ \overline{h} = \langle h \rangle = Ft \]
which is known to be correct for SOS models. (One might note that, strictly speaking, this
result is not `exact' for a finite system since the divergence term integrates to a
small but finite entity. Such `errors' are an inherent feature of the modelling of
finite systems by continuum differential equations). Thus, for SOS models invoking
`surface relaxation' we expect the generalisation of the equation of motion Eq (\ref{e2}), for
(pure) RD to be 
\begin{equation}
\frac{\partial h}{\partial t} = F + \frac{\partial G}{\partial x} + \eta
\label{e5}
\end {equation}
where $G$ is, at this stage, some unspecified function.
A computer model of $d'=1$ RD plus surface relaxation was evaluated by Family
\cite{family} many years ago. According to this author the resulting surface was 
self-affine, 
 i.e.\ was described by Eq (\ref{e1})  with the value of the parameters $\alpha ,~
\beta $ and $z$  being independent of the maximum number of lattice sites $l$ over
which a particle was allowed to move horizontally (a point we will  return to later).
Furthermore Family \cite{family} found that the scaling properties of the surface (or
equivalently, the values of $ \alpha , ~ \beta $ and $z$ ) were consistent with the
choice 
\begin{equation}
G = \nu  \frac{\partial h}{\partial x} 
\label{e6}
\end {equation}
with $\nu $ constant.

The resulting equation 
\begin{equation}
\frac{\partial h}{\partial t} = F+ \nu  \frac{\partial ^2h}{\partial x^2} + \eta
\label{e7}
\end {equation}
is well known in the literature and is referred to as the Edwards Wilkinson EW equation.
\cite{edwards} It is important to note exactly how the mapping of the horizontal
and vertical motion onto a differential operator occurs in Eq (\ref{e7}). The entity
$\nu \frac{\partial ^2h}{\partial x^2}$ (with $\nu $ positive) corresponds (as shown  
in Fig. \ref{fig_diff}) to a movement of particles from the top of a `hill' to the
bottom of it, i.e a flattening of the `hill'. In other words it is consistent with
movements both horizontally and vertically downwards. This manifests itself 
clearly in that model of Family \cite{family} in which all possible downwards moves of a
particle are allowed to occur, and the particle moves to the nearest
neighbour position only in the horizontal movement (i.e. $ l=1$). The form of the
associated surface is shown in figure \ref{fig_familysur}. Examination of the latter
shows that the `vertical' separation of nearest neighbour columns is also small for this
situation. Hence
the model leads self-consistently (as a result of the rules of movement) to a situation
where both small horizontal and downward vertical motion is the norm. Thus the situation were
all possible downward motion is permitted to occur with 100 \% probability is
replicated in the mathematics by the differential form $\nu \frac{\partial ^2h}{\partial
x^2} $.

This having been established we now return to the claim by Family \cite{family} that
the parameters $\alpha ,~ \beta $ and $ z$ are independent of the magnitude of $l$. To
see that this cannot possibly be true let us take a $d'=1$ surface of length $L$ and
allow the number of horizontal moves to be up to a maximum of $L$. For such a situation
it is clear that we will obtain layer by layer growth - the reason being the following.
During the first monolayer coverage `islands' will develop of size $L_i$
say. With increasing coverage it becomes increasingly probable that the next particle
added will be on one of these islands. However, if the particle is allowed $L$
horizontal moves, it will migrate readily to the edge of the island, go over the side and
then adhere to the rim of it, thus extending the size of the island.  There are several
important results to be deduced from these considerations. First, if we have a surface of
total length $ L $, then as long as we allow horizontal moves up to some maximum $l$
say, we may well find a series of surfaces belonging to the same universality class.
However this can only be true as long as $l$ is less than some fraction of the total
`length' $L$ of the surface. Once $l$ exceeds this fraction, an increasing number of islands
will coalesce and we will very quickly obtain layer by layer growth. In short, dependent
on the size of the system, universality classes will only be obtained if we permit a
limited number of horizontal movements only. The precise number is ill defined at this
stage,  but  typically in the literature this number is chosen to be one or two only. 

\subsection{Surface Relaxation Involving Vertical Motion Downwards Only}
We consider next vertical motion. To this end we will examine first what is usually referred to
in the literature as (pure) ballistic deposition (BD). In such a model one imagines the
particles being deposited vertically onto a substrate in a random sequence. The rule of
movement is to move vertically down until the particle meets another particle. The latter can be
either directly underneath or at the side of the moving particle. In either event
movement is then terminated. As is well known, such a model does have overhangs and
vacancies present in it. Let us pretend for the moment that we do not know the equation
of motion for the evolution of such a surface and consider instead a variant on this
model.  Namely (pure) BD plus surface relaxation. The specification of the latter is
the following. If there is another particle immediately beneath the one under consideration
then no movement occurs. On the other hand, if there is no particle immediately beneath
the one under consideration move the latter down until it comes into contact with
either the substrate or another particle which lies immediately below it. At this point
motion ceases. In other words we are looking at a model of BD plus surface relaxation
where the latter is defined to be vertical movement downwards. At this stage the extent
of the movement downwards is ill defined. It is obvious that such a model of BD plus
vertical movement downward gives us exactly the same end result as (pure) RD. Hence the
equation of motion for BD plus vertical movement downward (or pure RD) is Eq (\ref{e2}).
However, the interface width for such a situation increases without limit and hence,
after an infinite time, the width is infinite. Correspondingly the vertical movements
downward must be infinite. What this shows us is that when we define the rules of
movement for BD plus vertical relaxation, we have no idea of the extent of the vertical
movement. However, the model, when evolved self-consistently according to these same rules,
shows us that the motion in the vertical direction is unlimited. Furthermore, although
vertical movement is occurring, the equation of motion, i.e. Eq (\ref{e2}), does not
contain any differential operators, the structure itself is free of vacancies and the
surface is not self-affine.

\subsubsection{Surface Relaxation Involving Vertical Motion Upwards Only}
Let us consider next a variant on this model but this time involving upward motion. In
essence, the previous model started from (pure) BD allowed vertical motion downwards and
ended up with (pure) RD. We will now consider the converse,  i.e. we will start with
(pure) RD allow vertical motion upwards and end up with (pure) BD. Thus the rules of motion
are the following. Choose a random number and add one particle to the corresponding
lattice point. If the added particle has no neighbouring points occupied by particles to
a higher level, then leave it where it is. On the other hand, if the added particle
finds higher columns next to it, move it up the side of the highest such column until
it is level with the top most point. Again at this stage we have no idea of the extent
of the upward motion that will ultimately be involved in such a model. It is well known
of course that (pure) BD, which is in essence the model we are describing, leads to a
surface width which ultimately saturates. In turn this means that the upward motion is
in fact finite and determined by the surface width. Furthermore the associated equation
of motion for the surface is well known. However we deliberately refrain from writing it
down at this stage for reasons that will be apparent later. Suffice it to say that our
original starting point was (pure) RD for which the equation of motion is Eq (\ref{e2}).
We then included what turns out to be finite upward motion. The latter can, apparently,
be described by adding two differential operators to Eq (\ref{e2}). One of these turns out
to have the same form as that given in Eq (\ref{e7}). However the other cannot be written
as a divergence term i.e. it is a non-conservative term and is in fact represented by a
non-linear operator.

Another interesting aspect of these two cases is the following. At the outset it is not
apparent in either model just what the extent of the vertical motion is. However the
models themselves evolve in such a way that the downward vertical motion is unlimited in
extent whereas the upward vertical motion is limited. This shows that the rules of
movement, which appear to be the `mirror image' of one another,  lead to models which evolve in such a way that up-down symmetry in the
growth direction is broken,  i.e. motion downwards can occur to a significantly
greater extent than motion upwards.

\subsubsection{Surface Relaxation Involving Limited Vertical Motion Downwards Only}
We now wish to consider a third model which, as we will show, turns out to be a hybrid of
the previous two. When considering horizontal movement, we argued that the existence of
universality classes, associated with surface relaxation processes, was in fact a
consequence of the limited amount of horizontal movement that was permitted to occur. We
will now examine the corresponding  question of imposing a limited amount of vertical movement
downwards. Thus we begin with a model which initially starts from (pure) BD and then
allows a limited number of vertical moves downward (i.e.\ a maximum of 100 lattice
spacings say). Initially, when ``growing'' such a model on the computer, we will be dealing
with essentially pure RD. This is because the interface width in the early stages of
`growth' will be much less than 100 lattice spacings. Hence initially we will have a
SOS pure RD growth mode in which there are no vacancies present in the structure. For
this stage of the growth the governing equation will be Eq (\ref{e2}). However, as we
know, the interface width grows without limit for (pure) RD and a stage will eventually
be reached at which the interface width approaches and then exceeds 100 lattice
spacings. Once this occurs a situation will develop where some of the downward vertical
motion will leave vacancies in the structure, i.e. some of the downward motion will not
be SOS. For such a situation we would expect from our earlier considerations that the form of the
equation of motion would change and, by analogy with the previous case, we might
anticipate non-linear terms to appear in it. The hybrid nature of the present case 
from the earlier viewpoint of describing (pure) BD starting from (pure) RD is that only some of the particles having  higher
columns next to them are eventually moved up,  and even then they are only moved part of the
way up the column.

The essential point here is the following. We started from a model with well defined
rules of movement. In the early stages of growth, the resultant  structure was defect free (i.e. no
vacancies) and was described by a particular equation. However, as a self - consistent
result of the rules of movement, the surface eventually evolves according to a
different differential equation from the one describing the earlier stages of growth. 
During the second stage of growth defects (i.e. vacancies) do occur in the structure and
the resulting equation would be anticipated to contain non-linear terms. It is important
to stress that it is \underline{not} the case that the equation of motion is the same
throughout all the growth, but with a cross-over from one regime of dominance to another
regime. Rather different forms of equation are required to describe the evolution of the
different stages of growth. What is more, if we increased the maximum number of allowed
downwards vertical moves from 100 to say 1000, the time during which the first growth
stage was governed by pure RD would increase correspondingly. This shows clearly that the
``cross-over'' time between the two regimes is governed by the number of vertical moves
that are permitted to occur. Given that (pure) RD is not a self-affine surface, we can see
that in general whether such a surface (i.e. self-affine) is ever seen to evolve in a finite
time computer model can be crucially dependent on the nature and extent of the permitted
movements.

\subsection{Heuristic rules}
As a prelude to describing the proposed rules we  consider a simplified model which permits some
insight into this question. Consider first pure random deposition in one dimension and the
question of the change in the interface width following the deposition one monolayer. The
expression for the final height $\langle h \rangle _f$ reads
\begin{equation}
\langle h \rangle _f = \frac{1}{L}[\sum_{i} (h_i + 1 + \eta _i)] = \langle h \rangle _i + 1
\label{ea}
\end {equation}
where $\langle h \rangle _i$ is the initial average height and $L$ the `length' of the substrate.

Similarly 

\[ \langle h^2 \rangle _f = \frac{1}{L} [ \sum_{i} (h_i^2 + 2h_i +1) +2 \eta _i (h_i +1) + \eta
_i^2 ]  \]

i.e.
\begin{equation}
\langle h^2 \rangle _f \equiv \langle h^2 \rangle _i +2 \langle h \rangle _i + 1 + \langle 
\eta _i^2 \rangle
\label{eb}
\end {equation}

It follows from equations \ref{ea} and \ref{eb} that the final inteface width $w_f$ is related
to the initial inteface width $w_i$ by the relation 
\begin{equation}
w_f^2 = \langle h ^2 \rangle _f - \langle h \rangle _f^2 = w_i^2 +\langle \eta _i^2 \rangle
\label{ec}
\end {equation}

If we now imagine a relaxation process occuring involving the movement of particles downwards in
which a fraction $\alpha $ of the columns lose a particle and a fraction $\alpha $ of them gain
one, a similar analysis leads readily to the result

\begin{equation}
w_f^2 = w_i^2 + \langle \eta _i^2 \rangle + 2 \alpha + 2 \alpha [\langle h_i \rangle _g -
 \langle h_i \rangle _L ]
\label{ed}
\end {equation}
where $\langle h_i \rangle _g (\langle h_i \rangle _L )$ is the aveage height of the columns
which gain (lose) a particle. It follows from Eq. \ref{ed} that saturation will occur if the
following relation is obeyed.
\begin{equation}
\langle h_i \rangle _L - \langle h_i \rangle _g = \frac{1}{2 \alpha} \langle \eta _i^2 \rangle
+1
\label{ee}
\end {equation}
Since the noise term (i.e. $\langle \eta _i^2 \rangle$) is fixed it is clear that it will be
difficult, in general, to satisfy this equation on a layer by layer growth basis and that in
general large fluctuations (or oscillations) about the `equilibrum' width will occur in computer
simulations on finite size substrates. Another important feature that is apparent from
examination of Eq. \ref{ee}  is the following. If, within a class of allowed moves, we permit
only a fraction $\beta $ of any given move to occur, we would replace $\alpha $ by the fraction
$\alpha  \beta $. This means that the entity on the left hand side would increase progressively
as we decreased the value of $\beta $. In turn this implies that the corresponding interface
width would increase. Since allowed moves of a given class are presumed to be 
associated with a universality class of a particular type, this implies that the interface width
of the latter can be increased continously simply by reducing the probability of movement
uniformly for all allowed moves in the class (clearly the range of possible values of $\beta $,
although undefined at this stage), can not  be over the entire region $0 \leq \beta \leq 1$ since
e.g. $\beta = 0 $ gives the model involving no relaxation i.e. exactly how small $\beta $ can be
is not known at this stage. Put alternatively, this suggests that interface
width alone is a poor indicator of the universality class since the prefactor of the term
describing the time evolution of this width for such a class can be increased by simply
varying the probability of occurence of all allowed moves in the class.
It is revealing to pursue this argument further and to distinguish between the various types of
move contained within a given class of moves. This can be done on the basis of the co-ordination
number $v_i$ of the particle prior to movement. For example in case of $v_i = 1$ we could distinguish
between the fraction of particles $\alpha _1$ having a $v_i=1$ and sat on top of a column, from
the fraction $\alpha _2$ having a $v_i =1$ and sat next to a step edge. Yet again these are to
be distinguished from the fraction $\alpha _3$ of particles having a $v_i =2$. For generality we
will assume that the fraction of particles are $\alpha _1 $ up to $\alpha _n$. (In terms of the
previous notation ($\alpha = \alpha _1 +\alpha _2 + \ldots + \alpha _n $).
The generalisation of Eq. \ref{ed} can be readily shown to be

\[ w_f^2 = w_i^2 + \langle \eta _i^2 \rangle + 2 \alpha +2\{ \alpha _1 [ \langle h_i \rangle _g
 - \langle h_i \rangle _L \} + \]
\begin{equation}
 + \alpha _2 [ \langle h_i \rangle _{2g} - \langle h_i \rangle _{2L} ] +
\ldots + \alpha _n [ \langle h_i \rangle _{ng} - \langle h_i \rangle _{nL} ] \}
\label{ef}
\end {equation}

where, in an obvious notation $\langle h_i \rangle _{ng} (\langle h_i \rangle _{nL} ) $
is the average height of type $n$-columns that gain (lose) a particle. Once again the effect of reducing
all the $\alpha _i $ by the \underline{same} factor $\beta $ will give the immediate
generalisation of the result described earlier. However it is clear from the present equation
thath the functional form of the equation remains unchanged only when all the $\alpha _i$ are
reduced in this way. As opposed to this, if the $\alpha _i $ are all reduced by different 
factors $\beta _i$, then the functional form of the equation is changed. For such a situation it
is far from obvious whether the resultant class of moves belongs to the same universality class
as for $\beta = 1$ (or for a uniform value of $\beta $ ). In the extreme case that $\beta _i =
0$ for some of the moves, but exists for others, we would clearly anticipate a different
universality class. However, even for the latter situation, it is not clear whether increasing
the $\beta _i$ values which were previously identically zero to extremely small values would
change the universality class type,  i.e. the stability of the latter to small changes of
movement type is not presently known. These questions will be investigated in  subsequent
publications.

\subsection{Heuristic arguments for the differential operators appearing in the
stochastic differential equation}

To date, by considering simple ``growth'' mechanisms, we have argued that the question of
the nature of the evolving surface, and the precise form of the stochastic equation
describing its evolution, is determined essentially by the nature and extent of the
`horizontal' and `vertical' movements that are permitted to occur. In the present
section, based on our considerations in the previous one, we will present simple heuristic arguments for the form of the differential
operators to be associated with such movements.

Before proceeding to detail it is informative to consider the range of possibilities
that could occur for an evolving surface. Clearly the associated interface width could
either saturate or not. For the latter, there are then two possibilities - namely that
the associated stochastic growth equation contains differential operators or it does
not. If on the other hand the interface width does saturate it is almost certain that
the associated growth equation does contain differential operators. The question is
which ones?  

We form next a set of rules for which differential operators are to be associated with which
situation (see also Figure \ref{fig_diff} ). In the latter, following conventional wisdom, we have
considered four types of differential operators only namely 
$ (\frac{\partial h}{\partial x})^2,~
 \frac{\partial ^2h}{\partial x^2},~
\frac{\partial ^2}{\partial x^2} (\frac{\partial h}{\partial x})^2 $ and $  
\frac{\partial ^4h}{\partial x^4} $
(As should be clear from Fig \ref{fig_diff} the effect of such operators on a `symmetric
hill' is symmetric - hence as long as the prescribed rules of deposition and surface
relaxation possess this left - right symmetry one would expect such even order operators).

Our suggested rules (based simply on heuristic arguments and arising from a
consideration of the different types of interfaces that evolve from simple growth
models) are the following.

For surfaces whose interface width saturates;
\begin{itemize}
\item  Movement vertically downwards (further refinement of these rules are presented
later)
\begin{enumerate}
\item employ the operator $+ \nu \frac{\partial ^2}{\partial x ^2} $ if the rules of
movement include all possible downward motion 
\item employ the operator  $-\kappa \frac{\partial ^4}{\partial x ^4} $ if the rules of
movement exclude certain $v_i =2 $ particles.
\end{enumerate}
\item Movement vertically upwards (further refinement of these rules are presented
later)
\begin{enumerate}
\item employ the operator  $+ \gamma (\frac{\partial h}{\partial x })^2 $ if the rules of
movement involve the creation of vacancies in the structure and correspond to maximal 
movement (with 100 \% probability) of particles up the side of columns but do \underline{not} involve movement to the
top of the adjacent column (i.e. non SOS models).
\item employ the operator $ \lambda \frac{\partial ^2}{\partial x^2} (\frac{\partial
h}{\partial x})^2 $ if the rules of movement do not involve the creation of vacancies in
the structure and correspond to the movement of particles (with 100 \% probability) upward from the top of one
column to the top of an adjacent column (i.e. SOS models). Note that upwards motion can, on its
own, destabilize a surface, and for SOS relaxation needs to be conterbalanced by an equivalent
downwards motion `current'
\end{enumerate}
\end{itemize}

For surfaces whose interface width does not saturate there are at  least two distinct
possibilities. Either the growth equation does not contain differential operators (e.g.
pure RD or, as we will argue in a later paper, pure shading)
or it contains an instability. To see how the latter could arise consider the case of
pure RD plus (limited) surface relaxation (i.e. the deposition process modelled by Family.
\cite{family}) As described earlier, the evolution of this surface in $d'=1$ is described
by Eq (\ref{e7}) and  snapshots of this surface at successive stages of growth is shown in
figure \ref{fig_familysur}, i.e we have a smooth surface, in this instance,  in which the rules of movement
lead, self-consistently, to a succession of small horizontal and small vertical downward
motions.

Suppose we have a video film of the evolution of this surface as it runs forward
in time and (unknown to the viewer) we instantly stop the forward
running of this film and instead run it backwards. Mathematically the resulting unwinding of the
surface relaxation process corresponds to time reversal and is equivalent to
replacing $t$ by $-t$ in Eq (\ref{e7}). One consequence of the latter is that we would
replace  $\nu $ by $-\nu $ in Eq (\ref{e7}). Clearly this would now correspond to small
vertical upward motion (see figure \ref{fig_familysur} for a graphic illustration of this) 
Eventually such upward motion  would `undo' the effects of the previous surface
relaxation processes and we would in effect move back towards pure RD - i.e. the surface
would become unstable as far as saturating the interface width was concerned. In other
words, the existence of a term $ -\lambda \frac{\partial ^2h}{\partial x^2} $ 
 could
destabilize the surface growth. (This result can also be readily established via the
formal machinery of the group renormalisation approach.) It should be clear from these
considerations that reversing the sign of the four differential operators we discussed
earlier could also result in instabilities in the associated differential equation.
Hence one has to examine carefully the permitted movements in order to ensure that such
movements do not ultimately lead, via self-consistency, to terms in the associated
differential equation which make its solution unstable. An example of this is given
later.

\subsection{Comparison Of The Heuristic Rules With Computer Models Reported In The
Literature}

Most computer models reported in the literature permit the deposited particle one move in
the horizontal direction. Which direction the particle moves vertically is also determined
by the rules of movement. In certain cases both upward and downward motion is
permitted, whereas in other cases upward (or downward) motion only is permitted. For many
cases reported in the literature the rules governing surface relaxation (which are almost invariably of the
SOS type)
are not even stated in this simple form,  rather, in an effort to mimick the physics, the
rules are defined in terms of the coordination number $v_i$ of the particle to be moved.
In particlular, when $v_i \geq 2$ movement is not usually permitted unless
the coordination number increases - i.e. in those situations where the coordination  would
be the same at the end of the movement as it was prior  to movement (the so-called
`tie' situation) no movement is permitted. On the other hand where the coordination
number increases via the movement, the rule could either be 
\begin{enumerate}
\item[{\bf (a)}] move to the nearest neighbour site that increases the value of $v_i$
\item[or \ {\bf (b)}] move to the nearest neighbour site which gives the maximum value of $v_i$. 
\end{enumerate}
The subtlety of these rules is that, because of the topology of the surface, (a) and (b) define different
types of allowed moves in different dimensions. For example Kotrla et al. \cite{kotrla} have
shown that both (a) and (b) lead to vertical motion downwards in $d'=1$. However in $d'=2$ rule (a)
leads to predominantly downward motion whereas rule (b) leads to both downward and
appreciable upward motion. (Further clarification of this point in quantitative terms will be given in a subsequent paper.)

We will now consider various models from this viewpoint. For the well known $d'=1$ case
studied by Family \cite{family} the `tie' situation occurs frequently for $v_i=2$
 and the rules of
movement permit the particle to move down only  under these circumstances.
Correspondingly the resulting interface is smooth - hence we are led to the conclusion that
if all moves downward are permitted (including the tie situation) the resulting interface is
smooth and the associated relaxation downward is described by the operator  
$ \nu \frac{\partial ^2 h}{\partial x ^2} $. As opposed to this it is found that if, in
$d'=1$, moves downwards for the tie  situation when $v_i = 2$ are forbidden (Figures
\ref{fig_wolf} and \ref{fig_dassarma}) and downward motion only allowed
if the coordination number is increased, we obtain a rough interface (Fig
\ref{fig_wolfsur} whose properties are
described by the operator $- \kappa \frac{\partial ^4 h}{\partial x ^4} $.  
\cite{wolf3,sarma}
The generalisation of these same rules to $d'=2$ presents an interesting
situation in that for case (a) discussed above, this leads to predominantly downward motion only and is
hence described by the operator  $-\kappa \frac{\partial ^4 h}{\partial x ^4} $. 
\cite{wolf3,sarma} However for case (b), significant upward motion is also allowed which is essentially SOS in nature.
The coresponding interface (which is rough) will be described by an equation in which the
downward motion is represented by the operator  $-\kappa \frac{\partial ^4 h}{\partial x
^4} $ whereas the upward motion is decribed by
 $\lambda \frac{\partial ^2 }{\partial x ^2}   ( \frac{\partial  h}{\partial x} )^2 $.
This is in agreement with what is found in the computer modelling \cite{kotrla,sarmaghasis}.
Hence we here have a situation where the equation of motion, because
of the permitted rules of movement, is governed by the following equation
\begin{equation}
 \frac{\partial h}{\partial t} = F -\kappa \frac{\partial ^4 h}{\partial x ^4} + \eta 
\label{e8}
\end{equation}
in $d'=1$, but is described by the equation
\begin{equation}
  \frac{\partial h}{\partial t} = F -\kappa \frac{\partial ^4 h}{\partial x ^4}
+ \lambda \frac{\partial ^2 }{\partial x ^2}   ( \frac{\partial  h}{\partial x} )^2 + \eta
\label{e9}
\end{equation}
in $d'=2$. We would argue, in contrast with   statements in the literature \cite{kotrla} that this
is not an unusual situation in physical terms but is merely a consequence of the fact that
the permitted movements in $d'=1$ and $d'=2$
 are different. Equally we would argue\cite{barabasi} that it is not true that the
equation of motion is the same in $d'=1$ and $d'=2$ but that the crossover has not been seen in
$d'=1$.

Consider next the model of Lai and Das Sarma \cite{lai} These authors considered a similar
model in $d'=1$ to the one of Wolf and Villain \cite{wolf3}  and Das Sarma and Tamborenea.
\cite{sarma} However in the case of tie with $v_i =2 $ the particle was moved to the
nearest neighbour site with the smaller height difference (i.e. sometimes upwards and
sometimes downward motion occurred). For the case of a tie with $v_i=1$ the particle was
allowed to diffuse to higher bonding within a distance $l$ - i.e. for $d'=1$ downward motion
could also occur for this situation. Given that these rules are all SOS type and that
significant ammounts of 
upward and downward motion is occurring, we would expect that the resulting equation is the
fourth order non-linear equation appearing in Eq (\ref{e9}) . This is precisely what Lai and das Sarma
\cite{lai} found. 

In view of our earlier comments we would envisage an interesting situation developing
if, in the case of tie with $v_i =2$, the downward motion was forbidden and only the upward
motion was allowed. For such a situation we could anticipate an instability could develop in
the system since such upwards motion is not counterbalanced by an equivalent downwards motion
`current'. 
This is precisely what happens in the
model of Park et al. \cite{park} where no saturation of the interface width was found for
systems larger than a critical size.

As a final example we consider model 1 in $d'=2$ of Kotrla et. al. \cite{kotrla}. This model
is a straightforward generalisation of the Wolf and Villain \cite{wolf3} and Das Sarma and
Tamborenea \cite{sarma} model in that if there is a neighbouring site with a higher
value of $v_i$ the particle relaxes to it in  either the upward or the downward direction
In the case of a tie the particle remains where it is. It should be obvious from  our
previous discussion that once again we have SOS relaxation in the upward and the downward
direction and that the fourth order non linear equation i.e. Eq (\ref{e9})  will describe the evolution
of the surface. This is exactly what Kotrla et al. \cite{kotrla} found.

Given the success of our heuristic rules we could then invert the problem of predicting the
permitted movements to be associated with a given stochastic differential equation.
Consider for example the equation
\begin{equation}
 \frac{\partial h}{\partial t} = F + \nu \frac{\partial ^2 h}{\partial x ^2}  
+ \gamma (\frac{\partial  h}{\partial x }) ^2 +\eta ~ .
\label{e10}
\end{equation}

The terms $F+\eta $ imply that the initial deposition is SOS and is pure RD. The second
term represents movements horizontally and vertically downwards. Similarly the
term $(\frac{\partial  h}{\partial x }) ^2$ implies relaxation vertically upwards with the
particle being moved up the side of a column until it was (at most) placed level with
the top of it
(i.e. non SOS vertical motion which creates vacancies in the system). In summary, we would
expect that Eq (\ref{e10})  represented RD following by surface relaxation in which the process
involves both downward and upward motion. 
It comes as somewhat of a surprise therefore to find that Eq
(\ref{e10}) 
is apparently applicable in $d'=1$, to pure BD. If anything we might have expected that the
latter corresponds to (pure) RD followed by limited vertical movement upwards along the
side of  higher (neighbouring) columns - i.e. one might have expected (pure) BD
to have obeyed the equation
\[ \frac{\partial h}{\partial t} = F  
+ \gamma (\frac{\partial  h}{\partial x }) ^2 +\eta ~ .\]

A possible resolution of this paradox would be (as indicated by our previous reasoning)
that (pure) RD followed by surface relaxation in both the downward and upward  direction is
mathematically indistiguishable from pure BD. This problem will be addressed in future
publications. For the present we note that,  in graphical terms, what we are suggesting is 
that the situation depicted in Fig \ref{fig_towers} (a) (appropriate to `pure' BD) is ultimately
physically indistinguishable from that depicted in Fig  \ref{fig_towers} (b) i.e. RD plus  surface relaxation in
the downward and upward direction. Alternatively we could argue that we need both a downward and
an upward `current' of particles to obtain a stable interface.

\section{Conclusion}
In the present paper we have presented arguments to the effect that observation of universality classes
in computer simulations of surface growth is a consequence of the limited number of horizontal moves that
a given particle is permitted to make. (Conversely, if the particle was permitted to make any number
of moves up to a maximum value determined to be equal to the substrate size, then interface roughness
would never develop in the sense that growth would be of the layer by layer type). Given the limited
number of horizontal moves, the universality class is then determined by the nature and extent of the
vertical motion. This ultimately is controlled by the rules associated with the coordination number
$v_i$ taken in conjunction with the dimensionality $d'$ of the surface. In particular rules may be defined
which in effect allow for example
\begin{itemize}
\item[(a)] downward motion only in both $d'=1$ and $d'=2$
\item[or (b)] downward motion in $d'=1$ but both downward and upward motion in $d'=2$
\end{itemize}

If downward motion only is allowed and this is of the SOS type (i.e. a particle has another particle
beneath it at the beginning and end of the motion) then the degree of the differential operator 
$i.e.~ \frac{\partial ^2 h}{\partial x ^2}$ or $\frac{\partial ^4 h}{\partial x ^4} $ is determined by the
rules of motion associated with the tie situation. If motion downwards is allowed in the tie
situation, then the second order differential operator is appropriate since 
this corresponds to all moves downwards being permitted. On the other hand, if motion downwards is forbidden in the tie situation, then the fourth
order differential operator is applicable since such an operator is appropriate for the
situation where only a fraction of the moves that can occur are permitted to occur. 
Similarly if both upward and downward motion is allowed in the tie situation the fourth order equation is
again applicable.

In the case of upward motion the form of the differential operator will again be determined by the rules
of movement. If the upward motion is of the SOS type then it is decribed by the operator
$\frac{\partial ^2 }{\partial x ^2} (\frac{\partial  h}{\partial x })^2 $. However if it is non SOS type
(i.e. it leads to the creation of vacancies) it will be described by the operator 
$ (\frac{\partial  h}{\partial x })^2 $.

We have shown, in the body of the text, that these heuristic rules are consistent with many standard
results reported in the literature. Furthermore we have presented arguments to show that the nature and
extent of the permitted vertical motion can lead to situations where, for example, initially the growth
contains no vacancies and is described by a particular differential equation. However, eventually
vacancies will enter the structure (as a result of the rules of movement) and subsequently for this
situation, the growth is described by a different differential equation.

Further substantiation for the validity of these arguments will be presented in a series of subsequent
papers where we will also address the question of the apparent anomaly of the equation of motion for
pure BD, and the question of the universality class when the $\beta _i$ is chosen to have
different values for the different move types $i$.

\section*{Acknowledgements}
H. K. is supported by the Hans B\H{o}ckler Stiftung and is very grateful for his support.

%\centerline{\psfig{figure=diff.eps,width=12.0cm,angle=-90}}
%\begin{figure}
%\caption{ Differential Operators}
% \label{fig_diff}
%\end{figure}

\centerline{\psfig{figure=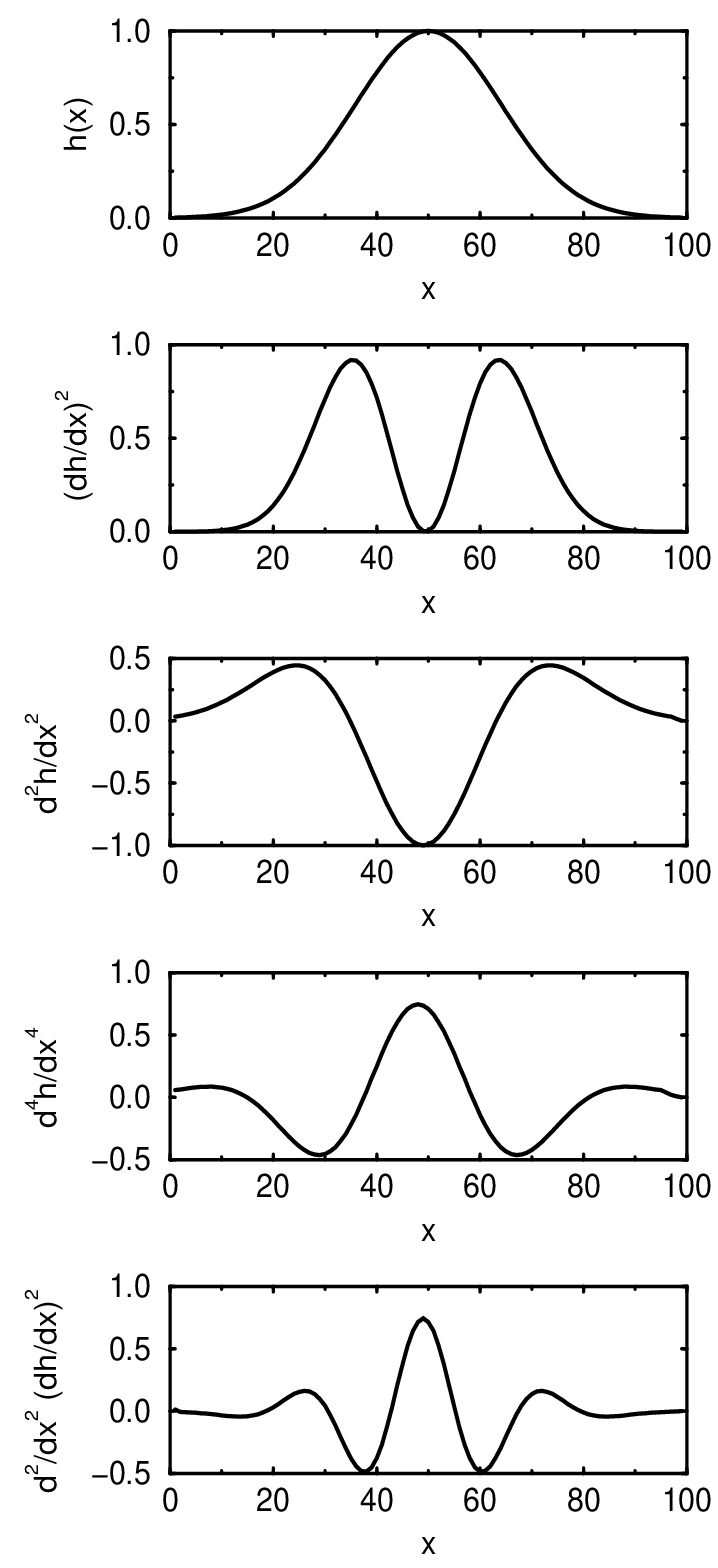,width=8cm,angle=180}}
\begin{figure}
\caption{ Differential Operators acting on the function $h(x)$  \\
$(+ \frac{\partial  h}{\partial x} )^2$ increases the particles at the side of the `hill' but does
not `transport' them to the top of it. \\
$+  \frac{\partial ^2 h }{\partial x ^2}$ removes particles form the top of the hill and redistributes
them to the sides and bottom of it.  \\
$- \frac{\partial ^4 h}{\partial x ^4}$ is similar to the previous case. \\
$+  \frac{\partial ^2 }{\partial x ^2}   ( \frac{\partial  h}{\partial x} )^2$ moves particles
from the base of the hill to the top of it. }
 \label{fig_diff}
\end{figure}

\centerline{\psfig{figure=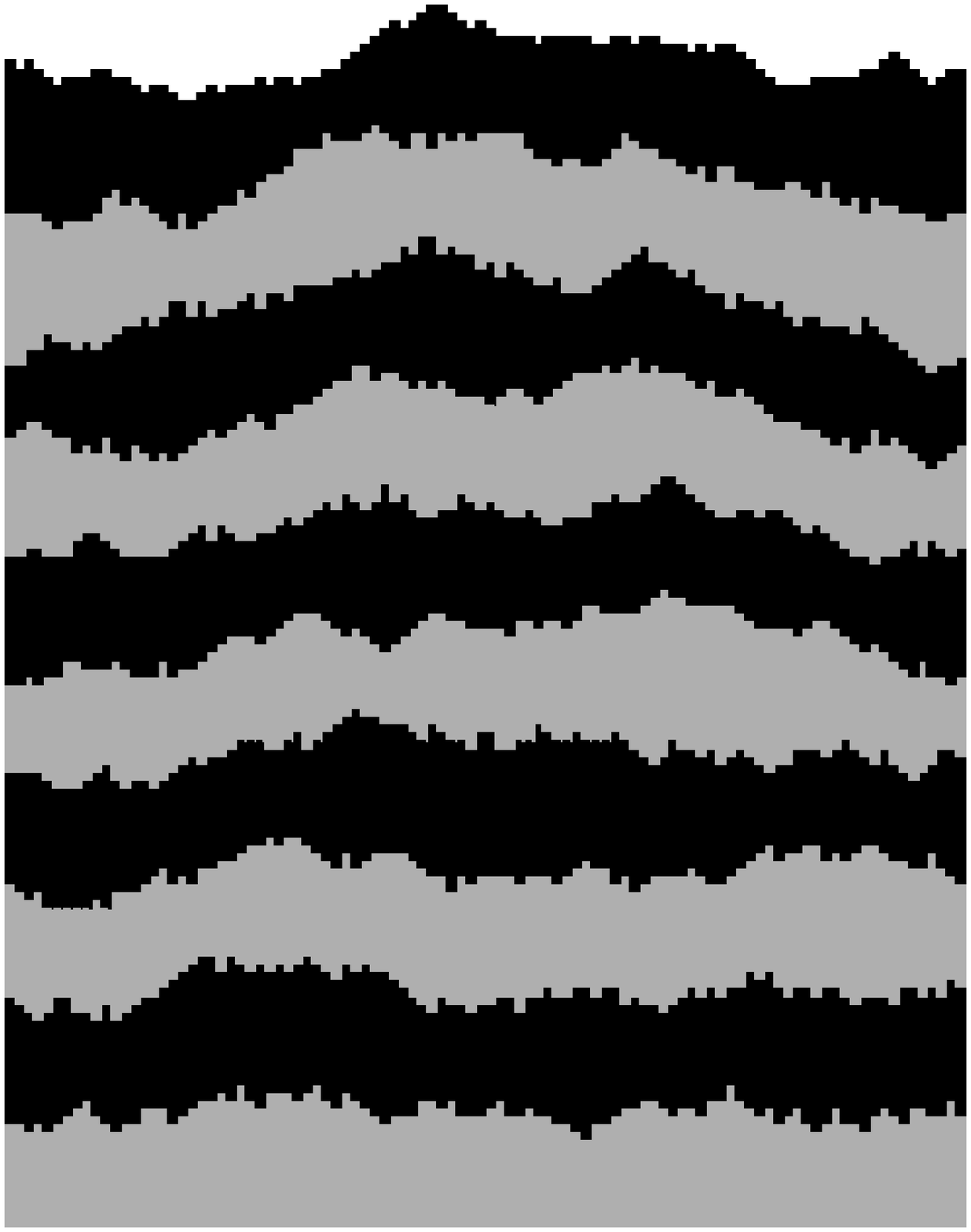,width=7.0cm,angle=0}}
\begin{figure}
\caption{Surface for RD plus relaxation after Family (taken from Ref. 1)  }
 \label{fig_familysur}
\end{figure}

\centerline{\psfig{figure=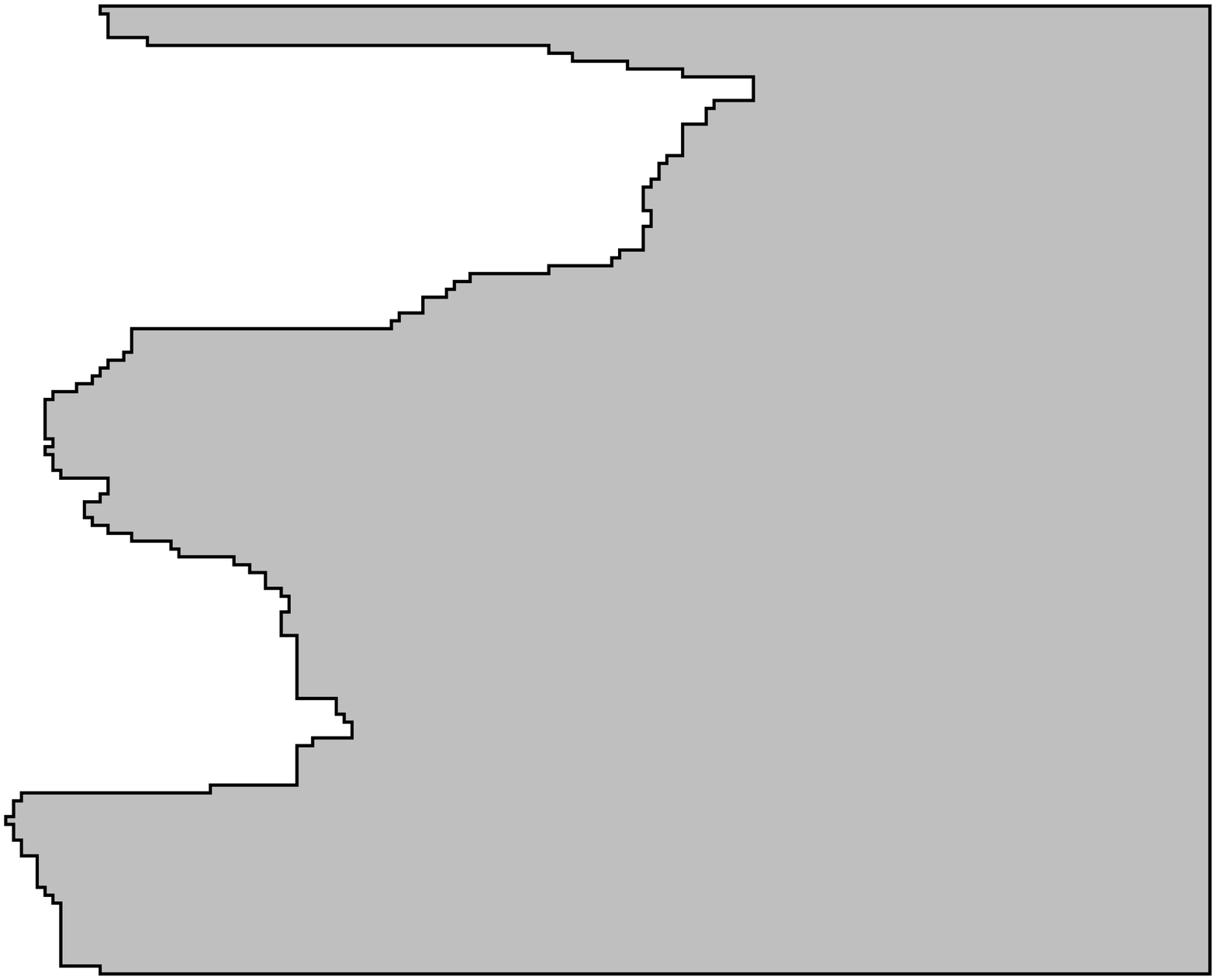,width=7.0cm,angle=-90}}
\begin{figure}
\caption{Surface of  the Wolf and Villain Modell (taken from Ref. 1) }
 \label{fig_wolfsur}
\end{figure}

\centerline{\psfig{figure=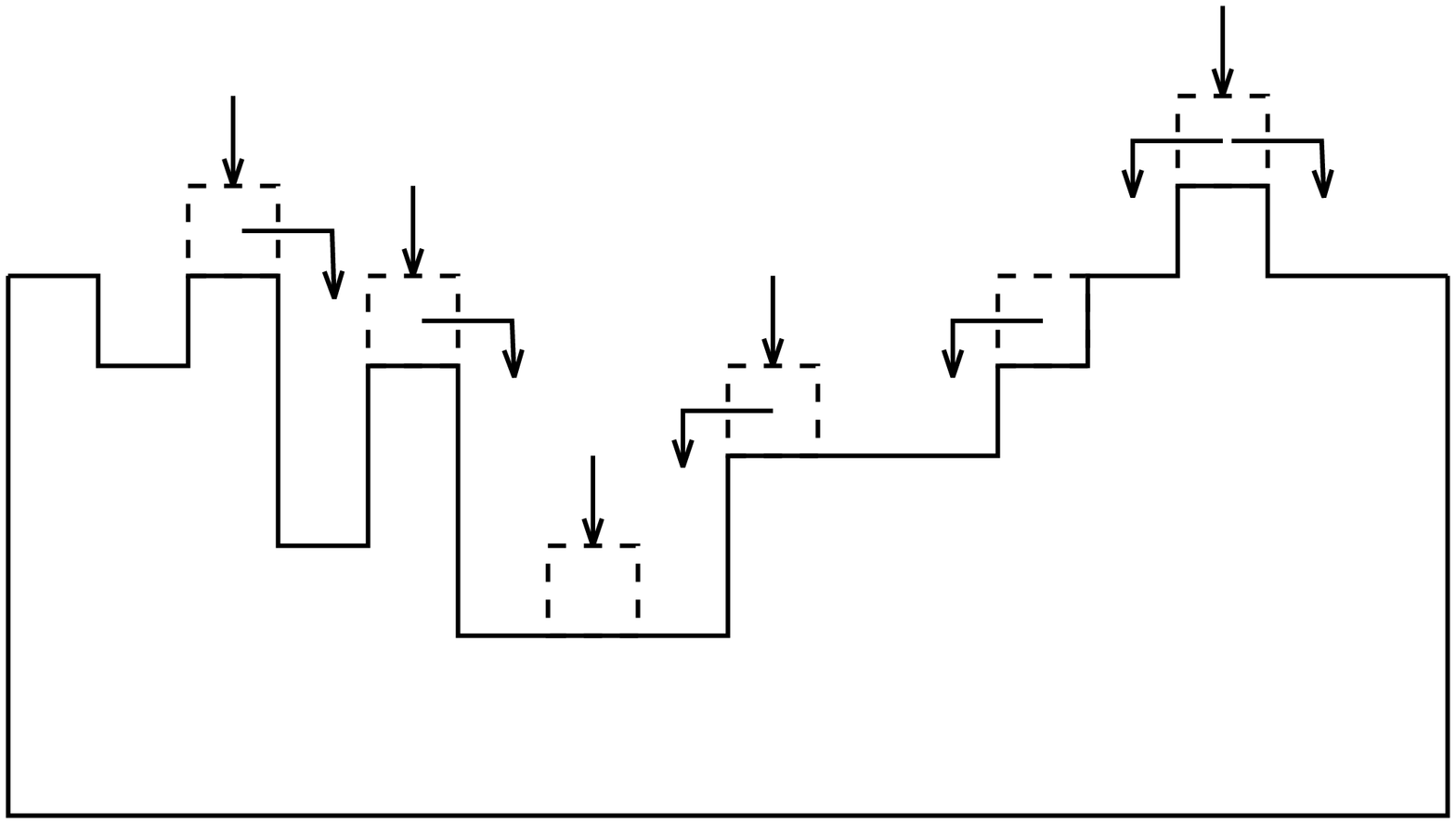,width=7.0cm,angle=-90}}
\begin{figure}
\caption{Growth rules of Family}
 \label{fig_family}
\end{figure}

\centerline{\psfig{figure=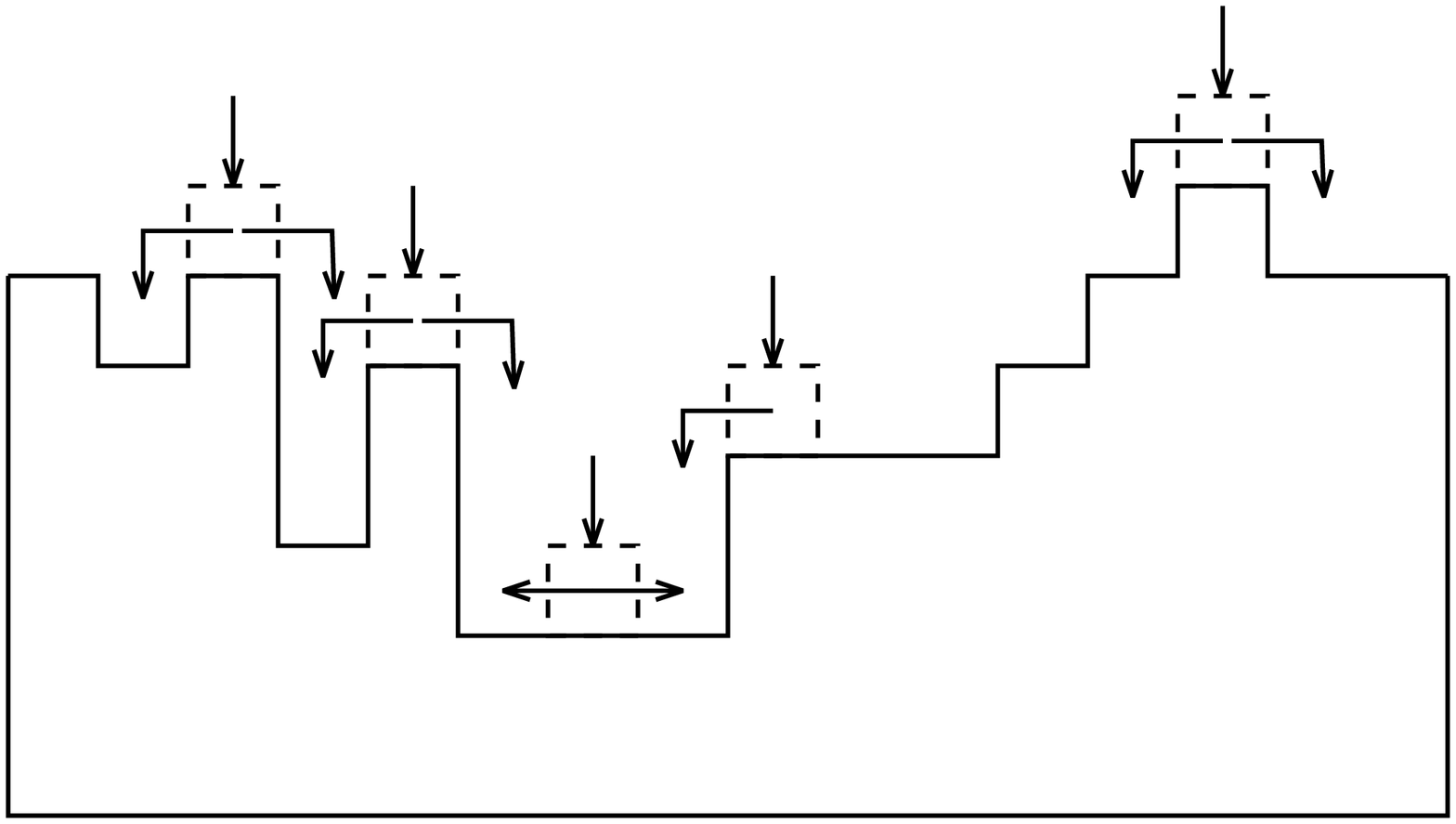,width=7.0cm,angle=-90}}
\begin{figure}
\caption{Growth rules of Das Sarma and Tamborenea}
 \label{fig_dassarma}
\end{figure}

\centerline{\psfig{figure=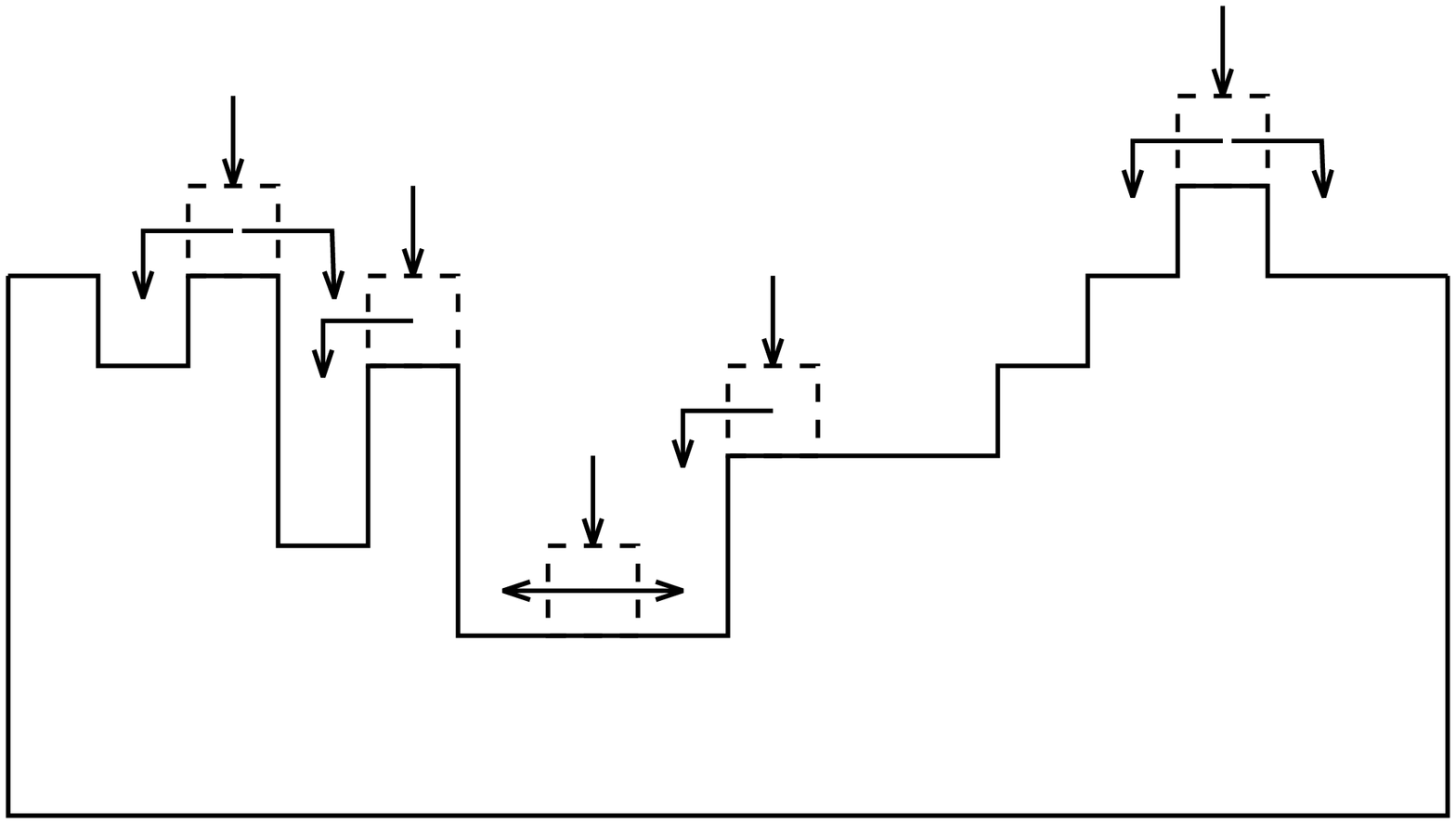,width=7.0cm,angle=-90}}
\begin{figure}
\caption{Growth rules of Wolf}
 \label{fig_wolf}
\end{figure}

\centerline{\psfig{figure=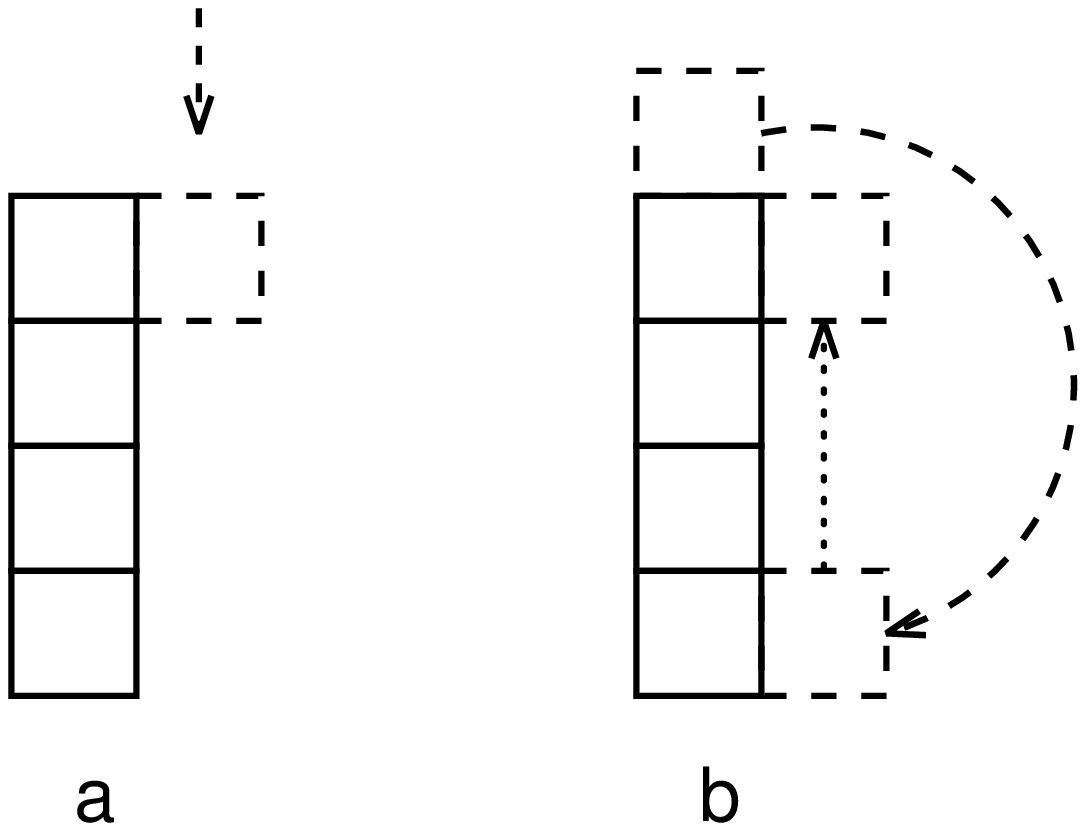,width=7.0cm,angle=-90}}
\begin{figure}
\caption{Different Rules With The Same Result}
 \label{fig_towers}
\end{figure}

\end{document}